\documentclass[12pt,preprint]{aastex}
\usepackage{epsfig}

\begin{document}

\title{Non-linear force-free field modeling of a solar active region
around the time of a major flare and coronal mass ejection}

\author{C.J.\ Schrijver\altaffilmark{1}, M.L.\ DeRosa\altaffilmark{1}, T.\ Metcalf\altaffilmark{2}, G.\ Barnes\altaffilmark{2},
B.\ Lites\altaffilmark{3}, T.\ Tarbell\altaffilmark{1}, J.\ McTiernan\altaffilmark{4}, G.\ Valori\altaffilmark{5},
T.\ Wiegelmann\altaffilmark{6}, M.S.\ Wheatland\altaffilmark{7}, T.\ Amari\altaffilmark{8}, G.\ Aulanier\altaffilmark{9},
P.\ D{\'e}moulin\altaffilmark{9}, M.\ Fuhrmann\altaffilmark{10}, K.\ Kusano\altaffilmark{11}, 
S.\ R{\'e}gnier\altaffilmark{12}, J.K.\ Thalmann\altaffilmark{6}}

\altaffiltext{1}{ Lockheed Martin Advanced Technology Center, Palo Alto, Ca., USA;}
\altaffiltext{2}{  Colorado Research Associates, Boulder, Co., USA; }
\altaffiltext{3}{  High Altitude Observatory, National Center for Atmospheric Research, Boulder, Co., USA; }
\altaffiltext{4}{  Space Sciences Laboratory, University of California, Berkeley, USA; }
\altaffiltext{5}{  Astrophysikalisches Institut Potsdam, Potsdam, Germany; }
\altaffiltext{6}{  Max-Planck Institut f{\"u}r Sonnensystemforschung, Katlenburg-Lindau Germany; }
\altaffiltext{7}{  School of Physics, University of Sydney, Australia; }
\altaffiltext{8}{  CNRS, Centre de Physique Theorique, Ecole Polytechnique, Palaiseau, France; }
\altaffiltext{9}{  LESIA, Observatoire de Paris, CNRS, UPMC, Universit{\'e} Paris Diderot, Meudon, France; }
\altaffiltext{10}{ Universit{\"a}t Potsdam, Institut f{\"u}r Physik, Potsdam, Germany;}
\altaffiltext{11}{  Earth Simulator Center, Japan Agency for Marine-Earth Science and Technology, Yokohama, Japan; }
\altaffiltext{12}{  School of Mathematics and Statistics, University of St. Andrews, Fife, UK.}

\begin{abstract}
Solar flares and coronal mass ejections are associated with rapid
changes in field connectivity and powered by the partial dissipation
of electrical currents in the solar atmosphere. A critical unanswered
question is whether the currents involved are induced by the motion of
pre-existing atmospheric magnetic flux subject to surface plasma
flows, or whether these currents are associated with the emergence of
flux from within the solar convective zone.  We address this problem
by applying state-of-the-art nonlinear force-free field (NLFFF)
modeling to the highest resolution and quality vector-magnetographic
data observed by the recently launched Hinode satellite on NOAA Active
Region 10930 around the time of a powerful X3.4 flare. We compute 14
NLFFF models with 4 different codes and a variety of boundary
conditions. We find that the model fields differ markedly in geometry,
energy content, and force-freeness. We discuss the relative merits of
these models in a general critique of present abilities to model the
coronal magnetic field based on surface vector field measurements. For
our application in particular, we find a fair agreement of the
best-fit model field with the observed coronal configuration, and
argue (1) that strong electrical currents emerge together with
magnetic flux preceding the flare, (2) that these currents are carried
in an ensemble of thin strands, (3) that the global pattern of these
currents and of field lines are compatible with a large-scale twisted
flux rope topology, and (4) that the $\sim 10^{32}$\,erg change in
energy associated with the coronal electrical currents suffices to
power the flare and its associated coronal mass ejection.
\end{abstract}

\keywords{Sun: activity, Sun: magnetic fields, Sun: flares, Sun: corona}

\section{Introduction}
Solar flares and coronal mass ejections derive their energy from
electrical currents that run through the solar outer atmosphere. There
is growing evidence that the strong electrical currents involved in
major flaring tend to emerge embedded within magnetic flux after being
generated within the solar convection zone, rather than being induced
by the displacement of pre-existing magnetic flux by plasma flows on
the surface
\citep[e.g.][]{leka+etal1996,wheatland2000,demoulin+etal2002,demoulin+etal2002a,falconer+etal2002,leka+barnes2003a,leka+barnes2003,schrijver+etal2005a,wiegelmann+etal2005,regnier+canfield2005,jing+etal2006,schrijver2007}. The
high-resolution vector-magnetographic capabilities of the recently
launched Hinode satellite, and the advances in computational
capabilities to model and analyze the atmospheric magnetic fields
based on these surface field measurements, should enable us to make
significant advances in addressing this problem.

Aside from flares and eruptive events, the magnetic field in the solar
corona evolves slowly as it responds to changes in the surface field,
implying that the electromagnetic Lorentz forces in this low-$\beta$
environment are relatively weak and that any electrical currents that
exist must be essentially parallel or anti-parallel to the magnetic
field wherever the field is not negligible.  The problem of
determining the coronal field and its embedded electrical currents
thus leads to the problem of reconstructing the 3D magnetic field from
the observed boundary conditions, without having to deal with the
effects of plasma forces on that field.

The vertical component of coronal electrical currents entering
the corona from below the photosphere can in principle
be deduced from the tangential components of the
vector-magnetic field at the base of the
corona. When combined with the vertical magnetic field component and
the condition that there are at most small
Lorentz forces (i.e., that the field is ``force free''), the resulting
model for the current-carrying coronal field is commonly referred to
as a nonlinear force-free (NLFF) field. Modeling such a field is in itself
a difficult problem that requires that a number of steps be taken successfully,
as outlined below \citep[see also, e.g.,][]{sakurai1989,mcclymont+etal1997,amari+etal1997}.

First, the measured polarization signals need to be inverted to form a
vector-magnetogram, which requires detailed models of radiative
transport of polarized light through the solar atmosphere.

Second, the procedure involves resolving an intrinsic 180$^\circ$
ambiguity in the components perpendicular to the line of sight, which
result from a degeneracy of the polarization properties. As the
electrical currents that penetrate the photosphere are carried by
compact flux tubes or potentially by fine structures within sunspots,
knowledge of the small-scale gradients in the field components
tangential to the solar surface is critical. Several procedures have
been developed to address this problem (see Section~2 for select
references), all of which involve a subjective choice, either on how
to deal with discontinuities interactively, or on what functional to
use in an automated iterative procedure.

Third, the Lorentz forces at the base of the corona (caused by
buoyancy forces and drag forces from surface plasma flows) must be
dealt with, because formally the assumption that currents and field
are collinear is valid only above the lower chromosphere
\citep{metcalf+etal95}, where the plasma $\beta$ rapidly lies well
below unity at least within the strong-field core of active
regions. This step, often referred to as preprocessing, also involves
subjective choices about how the field may be modified to remove net
forces and torques while smoothing and tilting the observed vector
field.  A parallel study by \citet{wiegelmann+etal2008} confirms the
expectation by \citet{metcalf+etal2007} based on a model test case
that successful preprocessing distorts the observed photospheric field to
an approximation of the chromospheric field.

Fourth, the NLFF field computation requires a numerical code to
determine a coronal field that is compatible with the observed
boundary condition.  This, too, involves choices about the iteration
scheme itself and about the application of boundary and initial
conditions. All of these affect the outcome, even in the case of
``perfect knowledge,'' as shown by the tests performed by
\citet{schrijver+etal2005b} and \citet{metcalf+etal2007}.

In a series of precursor studies, we have addressed the above set of
problems \citep[e.g.,][and references therein]{metcalf1994,wiegelmann+etal2005c,schrijver+etal2005b,amari+etal2005,metcalf+etal2007}. 
We here proceed with an application of the developed methodology to
state-of-the-art solar observations.

\section{Observations}
The Solar Optical Telescope (SOT) on board the Hinode spacecraft
\citep{kosugi+etl2007} observed AR\,10930 in the chromospheric 
Ca\,II\,H channel and in the near-photospheric G band, while also
obtaining magnetogram sequences for over a week with near-continuous
coverage \citep[see Figs.\,\ref{fig:1} and
\ref{fig:2}; the region's evolution is also 
described by, e.g.,][]{zhang+etal2007}.  These and other observations
show that AR\,10930 exhibited only B- and C-class flares from
2006/12/08 through 2006/12/13, when an X3.4 flare, peaking in soft
X-rays at 2:40\,UT, ended this relatively quiet period.  At the time
of the flare, the region was at 23$^\circ$\,west and 5$^\circ$\,south
of disk center, and thus well positioned for vector-magnetographic
observations.

If we look at this earliest phase in the flare, the very first
brightenings in the chromosphere (in Ca\,II\,H) are visible at
02:04\,UT over a pair of converging concentrations of opposite sign in
the line-of-sight magnetic field (at position A in
Fig.\,\ref{fig:2}d).  Twelve minutes later, three flare ribbons are
evident over opposite line-of-sight magnetic polarity
(Fig.\,\ref{fig:1}b): at position E and at the penumbral edges at
positions B and D in Fig.~\ref{fig:2}c).  Such flare ribbons are
commonly interpreted as the sites where energetic flare particles
impact the lower atmosphere, thereby identifying the photospheric end
points of the field lines on which these particles are accelerated
during the energization of the flare. Hinode/XRT observed an early
bright soft X-ray feature over the first chromospheric brightening
(near E in Fig.\,\ref{fig:2}c). The east-west bright X-ray ridge
(below the center in Fig.\,\ref{fig:1}a) straddles the emerging
flux ends on the northern early flare ribbon near position B in
Fig.\,\ref{fig:2}c and on the southern one near position D around
02:15\,UT. The overlying higher arched loops do not exhibit bright
ribbons until approximately 2:30\,UT (at positions C and near D in
Fig.\,\ref{fig:2}c); conversion of its excess energy into particle
kinetic energy apparently starts only some 25\,min after the first
impulsive energy conversion.

In the days leading up to the X3.4 flare, the main changes in
AR\,10930 comprised a strong eastward motion of the smaller, southern
sunspot relative to its larger neighbor to the north.
Flux emergence between the two spots, as well as in the area
west of that, continued strongly from the early hours of
2006/12/10 through the second half of 2006/12/14.  Towards the end of
that period, the southern spot was moving rapidly eastward, while
multiple ridges of both polarities showed up between the northern and
southern spots, some even forming interpenumbral connections. Snapshots
of four Hinode magnetograms taken with the Narrow-band Filter Imager (NFI)
are shown in Fig.~\ref{fig:2}. Note that the NFI signal is highly
non-linear within the spot umbrae, where the very strong field causes the
signal to fade back to zero; this does not happen in the spectrum-based
vector-magnetographic SP data 
(described below) that we use as input for our NLFFF methods. 
We show a selection of 
the NFI images here, because they are part of a movie (see below) with
magnetograms taken at a cadence of 
two minutes, whereas there is an interval of 8\,h between the SP maps
before and after the flare. 

Figure\,\ref{fig:2} shows only a small area around the sunspots at
4\,h intervals up to the start of the X3.4 flare. These panels suggest
relatively little change over time. The overall
appearance, in fact, changes so little
that we use the lower two panels to show the
detailed positions of the early flare ribbons and the modeled
electrical currents described in detail below. But the field is, in fact,
very dynamic: successive generations of ridges and concentrations of
flux form and disappear along the region between the spots and in the spots'
adjacent penumbrae. This evolution can be seen in an 8\,d movie in the
electronic addenda, with 2\,min between successive magnetograms (shown
rebinned to a 1,000 by 500 pixel movie at 1/4 of the full instrumental
resolution).\footnote{The movie is also available at \hbox{http://www.lmsal.com/$\sim$schryver/NLFFF/HinodeNFI$\_$X3.4.mov}. A smaller version covering a 2h\,h interval around the flare is also available.}

The evolution of the emerging field between the spots is characterized
by the frequent occurrence of opposite-polarity ridges, either next to
each other, offset along their length, or separated by a strip of
weaker line-of-sight field. Such nearly parallel strands are a
characteristic signature of emerging flux bundles that carry currents
along their core \citep[see, e.g., Fig.\,2 in ][for simulations of an
emerging flux rope]{magara2006}. Such fibril electrical
currents cause the magnetic field to spiral about the axis of a flux
rope, so that when this flux rope breaches the solar surface, two
ridges are observed where the spiraling field points upward and
downward in two mostly parallel ridges in close proximity on either
side of the rope's axis. We return to this point in the discussion of
the model field.

In addition to the filtergram sequences, the SOT Spectro-Polarimeter
(SP) obtained maps of the central regions of AR\,10930 before
and after the X3.4 flare. A pre-flare map was obtained between
20:30\,UT and 21:15\,UT on 2006/12/12, and a subsequent post-flare
map between 3:40\,UT and 4:40\,UT on the next day; both have
0.3\,arcsec pixels, and span 1024 steps with 512 pixels along the
north-south oriented slit.

These Hinode SP data were prepared by the standard ``SP$\_$PREP''
\citep{lites+etal2007a} available through SolarSoft, and the resulting
polarization spectra inverted to a vector magnetic field map
using an Unno-Rachkovsky inversion with a Milne-Eddington atmosphere
\citep{skumanich+lites1987,lites+skumanich1990,lites+etal1993}.  
Pixels with a net polarization below
the threshold required for full inversion, but still with measureable Stokes
$V$, are treated in the following manner: a longitudinal flux density
is derived \citep[see, for example,][]{lites+etal2007b}, and assuming that the
field in these regions is in fact radially directed, the observing angle
allows the azimuths and inclinations to be determined in the observer's frame.
In this manner, less than 0.5\%\ of the pixels in the map are undetermined
and discontinuities due to the thresholds are minimized.

The vector magnetic field maps are then subjected to an ambiguity
resolution approach which uses simulated annealing to minimize a
functional of the electric current density and divergence-free
condition \citep{metcalf1994}.  The algorithm applied includes the
enhancements described in \citet{metcalf+etal2006}, and was the
top-performing automated method amongst those compared in that same
study.  The ambiguity resolution appears to work successfully for most
of the field of view, and leaves only a few
very small patches where the transverse field changes
discontinuously. We cannot tell whether these patches are artifacts or
real, and consequently do not attempt to remove them from the
processed vector field. The vertical components of the SP
vector-magnetic field before and after the flare are shown in
Figs.\,\ref{fig:1}c and
\ref{fig:1}d, respectively. Maps of the vertical current density, $j_z$,
are shown in Figs.\,\ref{fig:1}e, and\,f; we note that the overall
currents are balanced to within 0.7\%\ in the field of view, while the
currents within each polarity reveal a net current between the field
polarities of approximately 20\%\ for positive $B_z$ and 15\%\ in the
negative polarity. The line-of-sight components of the central area as
measured by the NFI are shown enlarged in Fig.\,\ref{fig:2}.  We note
that the polarity patterns seen in the filter-based NFI data match
those in the spectrum-based SP vector magnetograms (the latter are
available as addenda to the electronic edition of this
paper\footnote{FITS files are also available at
http://www.lmsal.com/$\sim$schryver/NLFFF/; the file contents is described
in the FITS header comment field.}), as expected for subsonic
flows affecting fields observed at 120\,m\AA\ away from the line
center position at 6302\,\AA.

We embedded the Hinode/SP maps in a much larger, 
lower-resolution SOHO/MDI \citep{soho} line-of-sight magnetogram in order
to incorporate information on flux outside the SP
map, subject to the current-free (potential) approximation. 
In order to be able to apply the NLFF codes with
present computational resources, the data are rebinned $2\times 2$, to
0.63\,arcsec pixels. A Green's function potential field
\citep{metcalf+etal2007} is computed for the entire expanded area
to serve as initial condition, and as side- and upper-boundary
conditions (where applicable for the methods; see \citet{metcalf+etal2007}). 
The central area of $150 \times 150$\,Mm
($203\times 203$\,arcsec, or $320\times 320$ pixels) is extracted for
modeling, together with the corresponding potential-field cube over
256 vertical pixels.

\section{NLFFF modeling}
We apply the four NLFF field algorithms described in
\citet{schrijver+etal2005b} and, where modified, by 
\citet{metcalf+etal2007}: the weighted optimization algorithm by
\citet{wiegelmann2004}, the uniformly weighted optimization model as
implemented by McTiernan \citep{wheatland+etal2000}, the
magneto-frictional code by \citet{valori+etal2005}, and the
current-field interaction method by \citet[][and references
therein]{wheatland2006,wheatland2007}.  For the Wheatland method,
three solutions are computed with different boundary conditions on
vertical current: currents are chosen from regions with positive
vertical field $B_z$; from regions with negative vertical field; and
based on an average of those two.

Apart from working with the observed (disambiguated) vector fields, we
also apply
preprocessing to the lower boundary vector field. This 
removes net magnetic forces and torques which should not exist in
the model. The preprocessing is performed with a method devised by
\citet{metcalf+etal2007}, which leaves $B_z$ unchanged, and
one by \citet{wiegelmann+etal2005c}, which allows all field components
to change. Both of these are applied with and without spatial
smoothing. The characteristic value of the rms difference between 
observed and preprocessed values for $B_{x,y}$ are $\sim 15$\%\ of
the standard deviation in $B_{x,y}$.
A summary of these preprocessing  algorithms
is included in \citet{metcalf+etal2007}.

The work by \citet{metcalf+etal2007} showed that the additional step
of preprocessing the observed photospheric vector-magnetic data
resulted in a marked improvement in the agreement between the
resulting NLFFF extrapolations and their model reference field. They
argue that this
result likely stems from the fact that the preprocessed
photospheric field is a good approximation
of the corresponding chromospheric field, where the Lorentz forces
are much weaker. This is confirmed by \citet{wiegelmann+etal2008} in
a detailed evaluation of the
preprocessing process. As in the earlier study,
we find here that the best fits are obtained for preprocessed data.
We return to this topic in Section~4.

Using these algorithms, we obtain 14 distinct model fields,
summarized in Table~I. We use
two different measures to identify the model that best fits the
observed corona, and one
that identifies the most internally consistent force-free field; 
all to these identify the same model field as ``best.''

A subjective goodness-of-fit is provided visually by comparing the
TRACE and Hinode/XRT images with the computed field lines, and
assessing the match for five characteristic signatures of the field,
labeled in Fig.\,\ref{fig:1}a as follows: 1) the sheared arcade
between the spots, 2) the eastern arch of loops around the southern
spot, 3) the low, nearly horizontal field west of the southern spot,
4) the arcade high over that horizontal field, and 5) the absence of
shear around the emerging flux northwest of the northern spot. Table~I
shows the model fields ordered by the resulting metric $Q_{\rm m}$:
each good or poor correspondence contributes a bonus or penalty to the
metric of $+1$ or $-1$, respectively, while an ambiguous
correspondence is not weighed in the value of $Q_{\rm m}$.  Only the
Wheatland positive-field solution ($Wh^+_{pp}$) applied to a
preprocessed lower boundary (including spatial smoothing) successfully
reproduces all five of these features.\footnote{Renderings of seven of
the NLFFF models and of the potential-field extrapolation 
for the preprocessed pre-flare fields are shown in
the electronic addenda. These images are also available at
\hbox{http://www.lmsal.com/$\sim$schryver/NLFFF/}.}

A second, objective measure for the goodness of fit is based on
finding field lines in the models that best match a set
of identified coronal loops in the Hinode/XRT and TRACE coronal
images, and computing the deviation between these in projection
against the solar disk.  Coronal observations of AR\,10930 were made
both with Hinode's XRT using its thin-Be/open filter wheel setting,
and with TRACE \citep{traceinstrument} in its 195\,\AA\ passband, both
with 1\,arcsec pixels. In order to increase the signal-to-noise level,
we use the geometric mean of sets of exposures, as shown in
Fig.\,\ref{fig:1}a, for loop tracing. These loop traces are then
compared to 100 field lines computed for each of the model fields with
starting points distributed within the modeled volume along the line
of sight through the midpoint of each traced loop $i$.  For each of these
field lines, we compute the area contained between the corresponding
loop trace and the field line projected against the plane of the sky;
this area $A_i$ is bounded by line segments that connect the ends of
the traced loops to the nearest points on the projected field
lines. The field lines with lowest value for $A_i$ are selected as the
best-fit field lines for each traced loop. The model field with the
lowest total value $\Sigma_i \min(A_i)$ for the set of traced loops is
identified as the best model.  Formally, the best fitting model field
is again the $Wh^+_{pp}$ solution, but Fig.\,\ref{fig:1}a shows that
the correspondence between model field and corona is far from perfect
- we return to this issue in Section~4.

A final criterion is based on the consistency of the model field with
the properties of a truly force-free field: the $Wh^+_{pp}$ solution
also has the lowest residual Lorentz forces (as measured by a
current-weighted angle between magnetic field and electrical current),
while being among the solutions with the lowest average absolute
divergence of the model field (both of these metrics should equal zero for a
perfect field). Table~I lists the values of these metrics. Note that
the metrics for field divergence are naturally grouped by NLFFF
algorithm type.

In the present study, the Wheatland $Wh^+_{pp}$ model outperforms that of the
weighted-optimization algorithm by Wiegelmann that provided the best
solutions in our earlier two trial studies
\citep{schrijver+etal2005b,metcalf+etal2007}.
Based on this single application, we cannot identify why the
$Wh^+_{pp}$ model performs best. In fact, we note that the 
current-field iteration procedure does not converge to a fixed
field for any of the various boundary conditions, but rather oscillates.
The code was run for 20 iterations in each case, at which stage the
field is still changing at successive iterations (on average by 2\%\ 
in the vector norm), although the energy was found to vary only slowly
over the final iterations. We note that this was also seen for
an analysis of test fields for a similar Grad-Rubin-based method
by \citet{amari+etal2005}. Despite this, the $Wh^+_{pp}$ model
is still the best model field. We expect that it performs best among
the other Wheatland solutions owing to the following property:
The Wheatland code requires that no electrical currents leave the
volume. As the lower-boundary vector field over-determines the
solution, the code uses the boundary current information either where
$B_z$ is positive or where it is negative, and then only where the
field strength exceeds 5\%\ of the maximum value and where the
horizontal field changes its azimuth by less than 120$^\circ$ between
neighboring pixels (this criterion is intended to limit the spurious
currents associated with incorrect disambiguations of the
perpendicular field component). The field lines with $B_z >0$ primarily emanate
from the smaller, southern spot, which mostly close within the volume
without major problems. This allows the $Wh^+_{pp}$ solution to
achieve a highest-fidelity representation of the real coronal field,
in marked contrast to the $Wh^-_{\rm pp}$ solution for $B_z < 0$, in
which the field is highly distorted.

In a comparison with other methods, we note that the Wiegelmann and
McTiernan algorithms attempt to find a solution by minimizing a
functional that includes a non-zero divergence of the magnetic field
(and thus indirectly a source for electrical currents). Thus, magnetic
flux and electrical currents can originate within the model's volume,
in particular near the lower boundary where many of the currents of
interest run. Similarly, the Valori code can allow for higher values
in the field's divergence in the attempt of attain lower residual
Lorentz forces.  The Wheatland code makes use of a vector potential,
and thus has an intrinsically low divergence (compare the values in
columns 5 and 8 in Table~I, see also \citet{wheatland2007}). Whereas
this yields a model field that best meets the formal requirement of
being force free, one might have expected that the artificial sources
would provide other methods additional freedom to find a better
matching field. But it apparently does not. 

We defer further discussion of the relative performance of the various
methods to a later study, in which the effects of different base
vector-magnetic fields and instrumental fields of view need to be
compared for a much larger sample of observations which, at present,
does not yet exist. We can add one observation on the potential effect
of the field of view. The $Wh_{\rm pp}^+$ algorithm was also applied
to a vector-field footprint extended westward to include the
relatively small amount of flux in the leading part of the region that
appears to be involved in the post-eruption arcade. This extension
makes the footprint twice as wide, and required lowering the model
volume by a factor of two to maintain the demand on computer
memory. The $Wh_{\rm pp}^+$ pre-flare model field for this extended
footprint is very similar to the model field discussed above, both in
terms of field lines and in terms of electrical currents, although the
free energy is lower by some 5\%. We conclude that our inferences
discussed below about the pre-flare field configuration are not
affected significantly by our choice of the field of view. The
$Wh_{\rm pp}^+$ model for the extended post-flare configuration did
not converge; that aspect will require substantial further study.

\section{Discussion and conclusions}
The best-fit $Wh^+_{pp}$ model NLFF field suggests that, prior to the
flare, the strongest electrical currents run between the main spot
groups. These currents connect the penumbrae through filamentary
currents that arch up to $h\approx 12$\,Mm over 
low-lying current strands that lie within $h\approx 6$\,Mm over
the emerging field between the spots. Not surprisingly, the bulk of
the free energy (i.e., the energy in excess of that of the
minimum-energy state given by the purely potential field) is
associated with these electrical currents (Figs.\,\ref{fig:1}c,\,d,
and~\ref{fig:2}). The main current strands connect footpoints that are
well separated, and that traverse a substantial distance over 
and along the region of
emerging flux between the spots. They connect the emerged and emerging
flux adjacent to the southern spot to the umbra of the northern spot
(Fig.\,\ref{fig:2}c, and the top panels in Fig.\,\ref{fig:3}).

In the best-fit $Wh^+_{pp}$ model, a low-lying, compact structure of
current-carrying emerging flux lies below the main current arcade in
the region containing opposite-polarity ridges of vertical field (see
the perspective volume renderings in Fig.\,\ref{fig:3}a, and also
Fig.~\ref{fig:4}c). This is the site at which the energy release in the
X3.4 flare starts its impulsive phase (see Fig.~\ref{fig:2}c,\,d).

The flare appears to tap energy from most of the flux system that has
emerged over the preceding days: the currents in the relatively high
arching arcade are drastically weakened after the flare
(Fig.~\ref{fig:4}a,\,b), whereas the long fibril of concentrated
current low above the solar surface completely disappears
(Fig.~\ref{fig:4}c,\,e). We interpret the flare/CME as a cataclysmic
energy drain from a current-carrying flux rope that emerged from below
the photosphere. The complex field evolution with many mixed-polarity
ridges that is observed in the Hinode/NFI magnetogram sequence
suggests that the flux rope is either a bundle of smaller strands
wrapped about each other, or a single rope with internal structure
both in terms of field strength and current density. The 8\,h interval
between the SOT-SP vector magnetograms is far too long to address this
issue.

With the $Wh^+_{pp}$ NLFFF model in hand, what can we say about
the topology of the pre-eruptive magnetic field?
The field line plots in Fig.~\ref{fig:3} can be interpreted as 
a low altitude sheared arcade between the spots underneath an
essentially potential field that is nearly orthogonal to the arcade;
such a configuration has been described by, e.g., 
\citet{devore+antiochos2000}. When considering 
the counterclockwise rotation of the southern, positive-polarity
sunspot evident in the NFI movie (see the electronic addenda) one may
alternatively infer an overall twisting of the magnetic field, as
has been observed, for example, by TRACE for the corona over some rotating
sunspots \citep[see, e.g.,][]{brown+etal2002a}.  Several distinct 
conceptual models exist for 
such flux ropes. One of these is for ``included'' flux
ropes which have strong internal axial fields (that carry
at most  weak currents) separated from
an external field by a current layer (this concept was used in the
study by \citet{metcalf+etal2007}.  Another is for ``circuit'' flux
ropes that are produced by (thick or thin) net currents along their
axes surrounded by field lines that spiral about the current
\citep[e.g.,][]{titov+demoulin1999,torok+kliem2007}. 
Then there are ``twisted'' flux ropes, that either may emerge as such
\citep[e.g.,][]{fan+gibson2004,amari+etal2004} or may form from photospheric 
vortical motions
\citep[e.g.,][]{amari+etal1996,torok+kliem2003,aulanier+etal2005}. 
These should show a large-scale bipolar direct current pattern
surrounded by external return currents (i.e., currents with opposite
direction within the same magnetic polarity); the existence of such
patterns continues to be debated \citep[see][and references
therein]{wheatland2000}.

A potential discriminator for the applicability of any of these
concepts to AR\,10930 is the distribution of pre-eruptive photospheric
and coronal electric currents as deduced from the vector magnetograms
(Fig.~\ref{fig:1}e) and the NLFF field model (Fig.~\ref{fig:4}a, c, and\,e).
On the scale of the entire active region, the current pattern in
AR10930 bears a striking resemblance to that of ``twisted'' flux
ropes. Comparing the pre-flare column of Fig.\,\ref{fig:4}
with those of Figs.~9 and 10
in \citet{aulanier+etal2005}, we see (1) a swirling bipolar pattern
in $j_z$ in and immediately above the
photosphere, with 
concentrations at the sunspots, and arching around the spots to form 
parallel lanes of vertical
current located on both sides of the large-scale 
neutral line between the spots; (2) narrower and fainter return currents 
away from the main
neutral line on the edges of the current concentrations; 
(3) an elongated strong patch in the coronal currents
right at the neutral line with essentially horizontal currents
and field. In contrast, the current pattern of
AR\,10930 does not match that expected for the other types of flux
ropes: sheared arcades are associated with parallel lanes of direct
and return currents, ``included'' ropes mostly show a single current
shell whose photospheric footpoints result in narrow parallel lanes,
and ``circuit'' flux ropes show unidirectional currents only.

In this picture, the pre-eruptive compact current carrying structure,
at the location of the strong soft X-ray emission observed before and
during the flare, would simply trace the shortest twisted field lines
of a large-scale twisted flux rope. This is based on the expectation
that a given end-to-end twist applied to short, low field lines
results in stronger currents than when the same twist is applied to
longer, higher field lines
\citep{aulanier+etal2005}.

The interesting similarity of our model field to theoretical models of
twisted flux ropes, while consistent with the evolution seen in the
NFI magnetograms, is encouraging, but will require confirmation beyond
this single example. The relative roles of pre-existing twist below
the photosphere and any added twist during the emergence associated
with plasma flows will need to be assessed.  Moreover, the current pattern
shown in Fig.\,{fig:4} clearly shows a strong degree of filamentation,
which might have non-negligible consequences in terms of field
topology.  This filamentation may be a consequence of a process that
is conducive to generating smaller scales (including, e.g., the
convective collapse upon emergence), or may in fact hold the key to
another formation process for the overall configuration.

The best-fit $Wh^+_{pp}$ solutions for the pre- and post-flare
observations show a decrease in the free energy from 32\%\ of the
potential-field energy to a post-flare configuration with only 14\%\
in excess of the potential field model for the post-flare state
(Table~I). This corresponds to a drop in free energy of $3\times
10^{32}$\,ergs, even as the total field energy in the potential field
extrapolation shows an increase by $\sim 10^{32}$\,ergs due to the
continuing emergence of flux in the 8\,h interval between the SOT-SP
maps.  This decrease in free energy is adequate to power an X-class
flare (with energies of $10^{31-32}$\,ergs;
\citet{hudson1991,bleybel+etal2002}) associated with a coronal 
mass ejection \citep[with energies up to about
$10^{32}$\,ergs;][]{hundhausen97}.

We note that the continuing emergence of flux between the pre- and
post-flare SP maps not only causes the energy in the post-flare
potential field to be larger than the pre-flare potential field (see
last line in Table~I), but also causes most model fields to have
energy ratios $E/E_{\rm p,pre}$ that increase from before to after the
flare. Only the best-fit $Wh_{pp}^+$ model shows a drop in energy,
likely because its strong pre-flare currents lose more energy than the
flux emergence adds in this particular model field. Note that some
model fields have energies below that of the potential field; causes
for this anomaly were discussed for a test field by
\citet{metcalf+etal2007}.

This exercise illustrates several problematic issues with the NLFFF
extrapolation process. First, the ``preprocessing'' of the observed
vector field yields a marked improvement in the quality of most of the
NLFF field models.  The preprocessing is one way of reducing the
effect of the Lorentz forces acting in the photosphere; further
studies of how best to deal with these forces should be undertaken
\citep[along the lines of that for a model field by][]{wiegelmann+etal2008}. 
Second, although provided with the same boundary conditions, only one
among seven model fields based on preprocessed boundary conditions
matches the observed coronal field acceptably by visual inspection
based on a limited set of key features (as measured by $Q_{\rm m}$ in
Table~I).  This reveals the sensitivity to details in model
implementation and boundary and initial conditions also seen in our
two earlier studies \citep{schrijver+etal2005b,metcalf+etal2007}.  An
integral part of the latter problem is, of course, the resolution of
the 180$^\circ$ ambiguity intrinsic to the measurement of the field
component perpendicular to the line of sight \citep[see the tests of
various methods based on known model fields described
by][]{metcalf+etal2006,li+etal2007}.  It is clear that a systematic
study of these differences between the algorithms ranging from the
observations to the final NLFF field models, and their relative merits
and problems, is needed to identify one or a
few of the most successful strategies.

And, finally, we note that even the best-fit $Wh^+_{pp}$ model
provides a rather poor match to the observed coronal loops. This may
be a consequence of the global nature of such loops: even if the field
locally would be modeled to within a few degrees, the integration
along the path of a field line could still lead to a markedly
different path relative to those in the true field and to observed
coronal loops. On the other hand, we note that the
poor match may be a consequence of an intrinsic problem of tracing
field lines by using coronal brightenings: \citet{aulanier+etal2005}, for
example, argue that  coronal S-shaped structures associated with
twisted flux ropes may not trace individual field lines, but rather
are formed by an ensemble of partially outlined loops subject to 
line-of-sight integration.

Although we can measure the divergence of the field line
relative to the observed loops, there is unfortunately no known method
that quantifies how significant such path differences are in terms of
the field's total energy or helicity. Both of the latter quantities
are of interest to, say, space weather forecasters, but we have yet to
learn how the comparison of observed loops to model field lines can
be used to improve estimates of energy and helicity. 

On the positive side, (1) the best-fit $Wh^+_{pp}$ model field
compares relatively well with the observed corona, and (2) the energy
estimates and the pre-to-post flare energy difference suffices to
power the flare. Consequently, we think that the best available
vector-magnetographic data and NLFFF modeling techniques support our
main finding:

We conclude that the filamentary electrical currents that emerge with
the magnetic flux between the main sunspots in AR\,10930 over a period
of up to several days (1) carry enough energy to power the observed
major X3.4 flare and associated coronal mass ejection, (2) are
involved in earliest impulsive phase of the flare, and (3) show a
substantial decrease in magnitude even as their associated field lines
connect over shorter distances when comparing pre- and post-flare
states. Thus, our application of nonlinear force-free field modeling
to state-of-the art vector-field data on a complex active region
provides strong evidence in support of the growing notion that major
solar flares are directly associated with the energy carried by
electrical currents that emerge from below the solar surface, and
is suggestive, at least in this case, of an emerging twisted flux rope.

\acknowledgements
Tom Metcalf died in a skiing accident before this manuscript was
completed; we dedicate it to the memory of his friendship and
collegiality.  Hinode is a Japanese mission developed and launched by
ISAS/JAXA, collaborating with NOAJ as domestic partner, NASA (USA) and
STFC (UK) as international partners.  CJS, MLD, TRM, and GB were
supported by by Lockheed Martin Independent Research Funds.  The work
of TW was supported by DLR grant 50 OC 0501 and that of JKT by DFG
grant WI 3211/1-1.  GV was supported by DFG grant HO1424/9-1.  GA and
PD gratefully acknowledge financial support by the European Commission
through the SOLAIRE Network (MTRN-CT-2006-035484).  MSW acknowledges
generous support provided by the Observatoire de Paris enabling travel
to the workshop.  The team thanks the CIAS of Observatoire de Paris
for hosting the NLFFF4 workshop.


\clearpage
\begin{table}
\caption{Metrics for the field extrapolations, in order of quality Q
based on the visual correspondence to the coronal pre-flare image.}
\begin{tabular}{l|r|lcr|lcr}\hline
\label{table:metrics}
&&\multicolumn{3}{c}{pre-flare: 2006/12/12} & \multicolumn{3}{c}{post-flare: 2006/12/13} \\  
Model\tablenotemark{1}  &$Q_{\rm m}$ \tablenotemark{2}& $E/E_{\rm p,pre}$ \tablenotemark{3} & ${\rm CW}\sin$ \tablenotemark{4}& $\langle |f_i|\rangle\times10^8$ \tablenotemark{5}& 
         $E/E_{\rm p,pre}$& ${\rm CW}\sin$& $\langle |f_i|\rangle\times10^8$\\
\hline

1: $Wh^+_{pp}$&5& 1.32& 0.24& 3.6 & 1.19& 0.18& 2.0\\

2: $Wh^+_{np}$&3& 1.10& 0.27& 3.9 & 1.23& 0.27& 4.6\\
3: $Wie_{wp}$&3& 1.09& 0.35& 19. & 1.18& 0.32& 13.\\
4: $Val_{pp}$&3& 1.10& 0.28& 230.& 1.27& 0.31& 190.\\

5: $Wh^0_{pp}$&2& 1.04& 0.28& 3.0 & 1.53& 0.27& 3.7\\
6: $Wie_{ns}$&2& 1.04& 0.43& 22. & 1.13& 0.39& 30.\\
7: $Val_{np}$&2& 0.88& 0.29& 220.& 0.99& 0.34& 210.\\

8: $Wie_{np}$&1& 0.95& 0.43& 24. & 1.04& 0.39& 27.\\

9: $Wie_{pp}$&0& 1.05& 0.44& 18. & 1.15& 0.39& 21.\\
10: $McT_{pp}$&0& 1.01& 0.61& 29. & 1.07& 0.59& 25.\\

11: $Wh^0_{np}$&-1& 1.03& 0.27& 2.5 & 1.12& 0.23& 2.6\\
12: $Wh^-_{np}$&-1& 1.04& 0.25& 2.9 & 1.11& 0.24& 2.9\\
13: $Wh^-_{pp}$&-1& 1.05& 0.27& 3.2 & 1.16& 0.19& 2.2\\

14: $McT_{np}$&-2& 0.95& 0.64& 26. & 1.00& 0.61& 24.\\

15: pot'l &-3& 1&   - & 0.8 & 1.04&    -& 0.8\\
\hline
\end{tabular}
\tablenotetext{1}{Models: Wh: Wheatland, Wie: Wiegelmann, Val: Valori, McT:
McTiernan; $+,-,0$: based on positive or negative $B_z$, or both,
respectively; np: no preprocessing; ns: preprocessed without
smoothing, pp: preprocessing including smoothing; wp: Wiegelmann's
preprocessing and smoothing.  }
\tablenotetext{2}{Quality of fit by visual
inspection for five features: a good/poor correspondence for each
feature adds $+1,-1$, respectively, to the total value; an ambiguous
correspondence adds $0$.  }
\tablenotetext{3}{Energy, relative to the energy in the
pre-flare potential field model. }
\tablenotetext{4}{Current-weighted value of
$\sin{\theta}$, where $\theta$ is the angle between the electrical
current and the magnetic field in the model solution.  }
\tablenotetext{5}{The
unsigned mean over all pixels $i$ in the comparison volume of the
absolute fractional flux change $|f_i| = {{|(\nabla\cdot {\bf
B})_i}|/{(6|{\bf B}|_i/\Delta x) }}$, where $\Delta x$ is the grid
spacing
\citep[compare][]{wheatland+etal2000}.}
\end{table}
\clearpage

\figcaption{The chromosphere, corona, and magnetic field of NOAA
AR\,10930. Panel {\em a} shows the $320 \times 320$ pixel footprint of
the full model volume for the NLFFF codes (itself surrounded by a much
larger skirt of a line-of-sight magnetic map).  The largest square is
the $224\times 224$ pixel area (with sides of 101\,Mm) used for the
energy estimates, and shown in panels {\em b--d} and in
Figs.~\ref{fig:3} and\,\ref{fig:4}. The smaller square shows the
footprint shown in Fig.~\ref{fig:2}. Panels: {\em a)} time-averaged
Hinode/XRT soft X-ray image. Individual loops traced on this image or
on a TRACE 195\,\AA\ image (not shown here) are represented in green;
the best-matching model field lines for the $Wh^+_{\rm pp}$ model
field are shown in purple.  Numbers identify the characteristic
signatures in the field that were used in the subjective assessment of
the goodness-of-fit, as discussed in Section~3.  {\em b)}
Chromospheric Ca\,II\,H image for the onset phase of the flare.  {\em
c)} Pre-flare Hinode/SOT-SP $B_z$ on 2006/12/12 around
21\,UT. Contours show the vertically integrated energy density in the
$Wh^+_{\rm pp}$ field minus that of the potential field; the white
contour lies near the maximum value and the green contour at half
that.  {\em d)} Post-flare Hinode/SOT-SP $B_z$ on 2006/12/13 around
4\,UT. Contours as in panel {\em b}.  {\em e, f)} Maps of the vertical
current density $j_z$ corresponding to the pre- and post-flare maps
shown in panels {\em c} and {\em d}, respectively. The color scale for
the lower four panels runs from blue (negative) to red (positive),
saturating into black or white, respectively. \label{fig:1}}

\figcaption{Time series of magnetograms showing the evolution of
the line-of-sight magnetic field as observed by the HINODE Solar
Optical Telescope with the Narrow-Band Filter Imager (NFI) at 4\,h
intervals prior to the X3.4 flare.  Coordinates (with north up and
-~by solar-physics convention~- west towards the right) are in pixels
of 0.16\,arcsec each; the area shown covers the central 10\%\ of the
footprint of the NLFF field model volume (compare the smallest square
in Fig.\,1a to its full field of view). Note that the NFI magnetograph
signal is non-monotonic, disappearing in the umbrae of the two spots.
Panel {\em d} shows in red the brightest segments of the flare ribbons
seen in the Ca\,II\,H channel at 02:16:39\,UT (cf.,
Fig.\,\ref{fig:1}d); it also outlines the brightest quiescent kernel
in soft X-rays seen by the Hinode X-ray telescope at the same time
(green contour) and the brightest coronal structure (black contour;
cf., Fig.\,\ref{fig:1}a). The latter two are repeated in panel {\em
c}, taken close to the time of the pre-flare vector-magnetogram
obtained by the spectro-polarimeter. The magnetic information in that
panel has been replaced by the line-of-sight integral of the
vector norm of the 
electrical currents (cf., Fig.~\ref{fig:4}a) within the blue contour,
which is where these currents are strongest.\label{fig:2}}

\figcaption{Visualizations of the magnetic field over NOAA
Active Region 10930 before (top) and after (bottom) the X3.4 flare,
shown against the corresponding map of $B_z$. Sample field lines
outline the field; white field lines close within the NLFF model
volume, while colored field lines (purple or green for the two
polarities of $B_z$ at their base) leave that volume to connect to
more distant regions. The rendered volumes (red) show where the
electrical current densities are highest, using the same threshold
level in both panels (cf., Fig.\,\ref{fig:2}c). The compact, low
current system below the large, high-arching currents in the top panel
corresponds to the site (position A in Fig.\,\ref{fig:2}d) of the
initial brightenings of the X3.4 flare and associated coronal mass
ejection.\label{fig:3}}

\figcaption{{\em a,\,b)} Vertically integrated electrical
currents in the $Wh^+_{\rm pp}$ pre-flare (left) and post-flare
(right) model fields, on the same color scale.  {\em c,\,d)} As panels
{\em a,\,b}, but showing the horizontal currents integrated over the
lowest 6,100\,km (12 pixels), using the same grey scale.  {\em e,\,f)}
As panels {\em c,\,d} for $j_z$, with white/black for
positive/negative $j_z$, respectively.\label{fig:4}}

\clearpage

\begin{figure}
\epsscale{.8}
\plotone{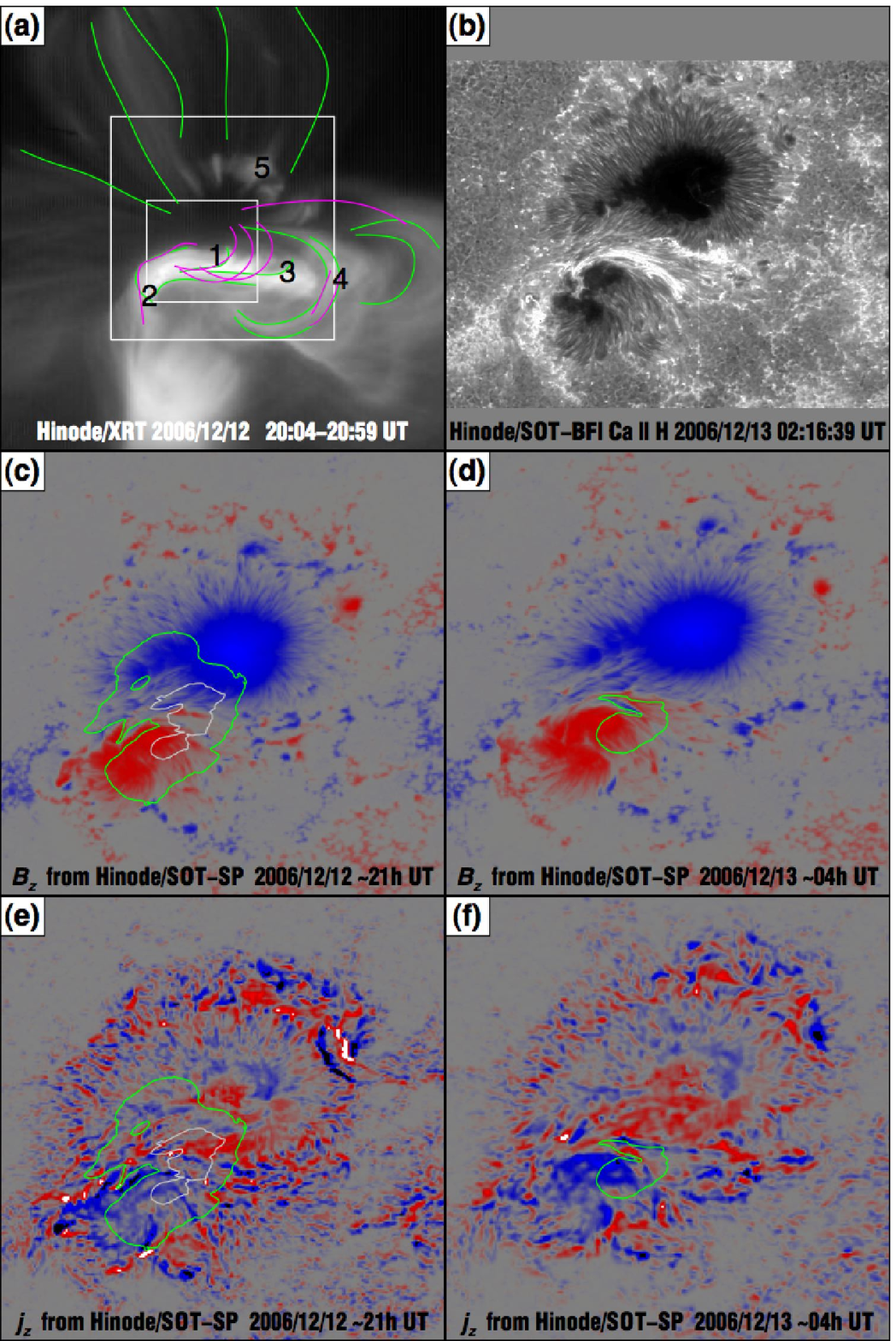}
\epsscale{1}
\centerline{f1.eps}
\end{figure}
\clearpage

\begin{figure}
\plotone{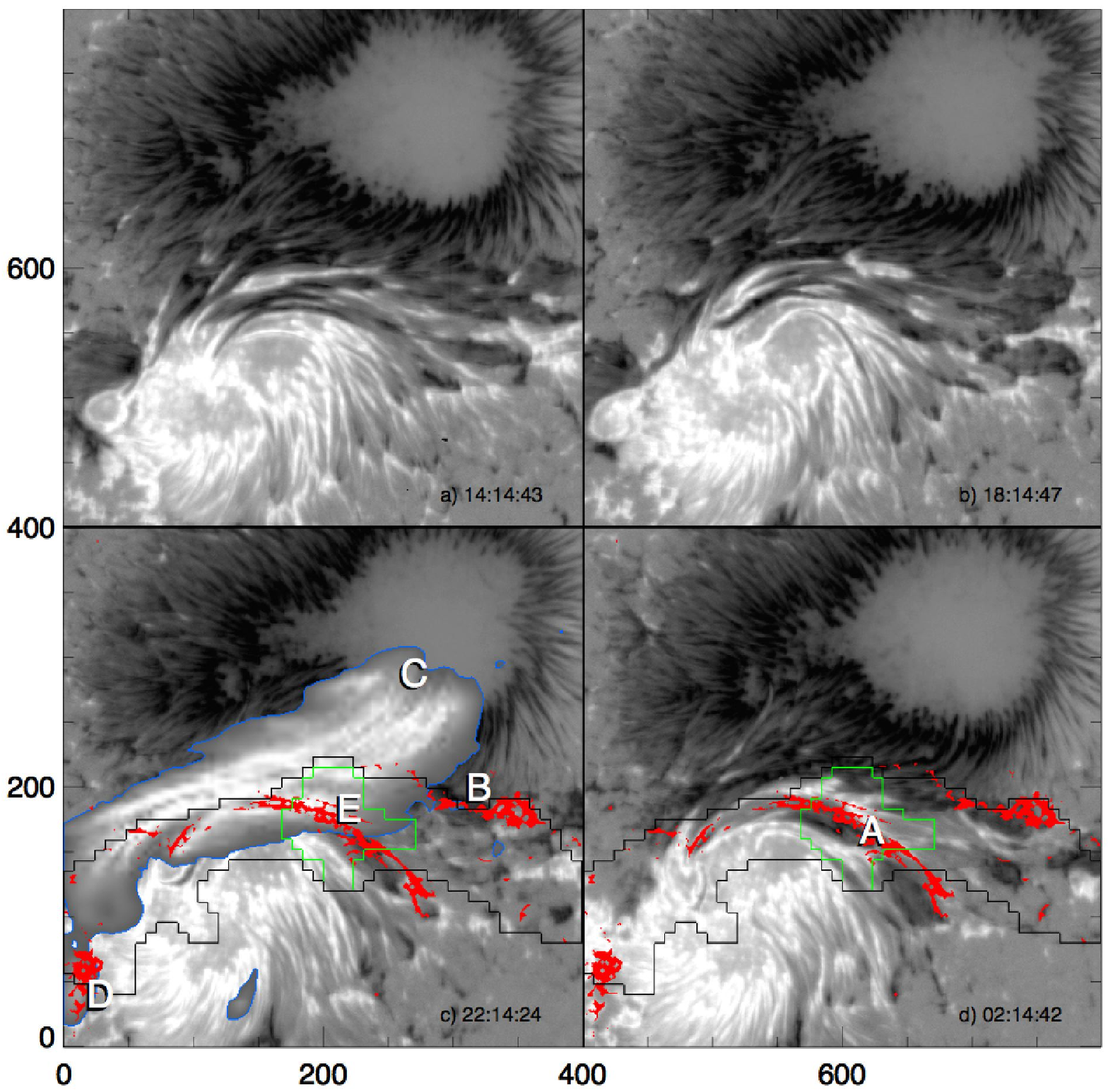}
\centerline{f2.ps}
\end{figure}
\clearpage

\begin{figure}
\centerline{\psfig{figure=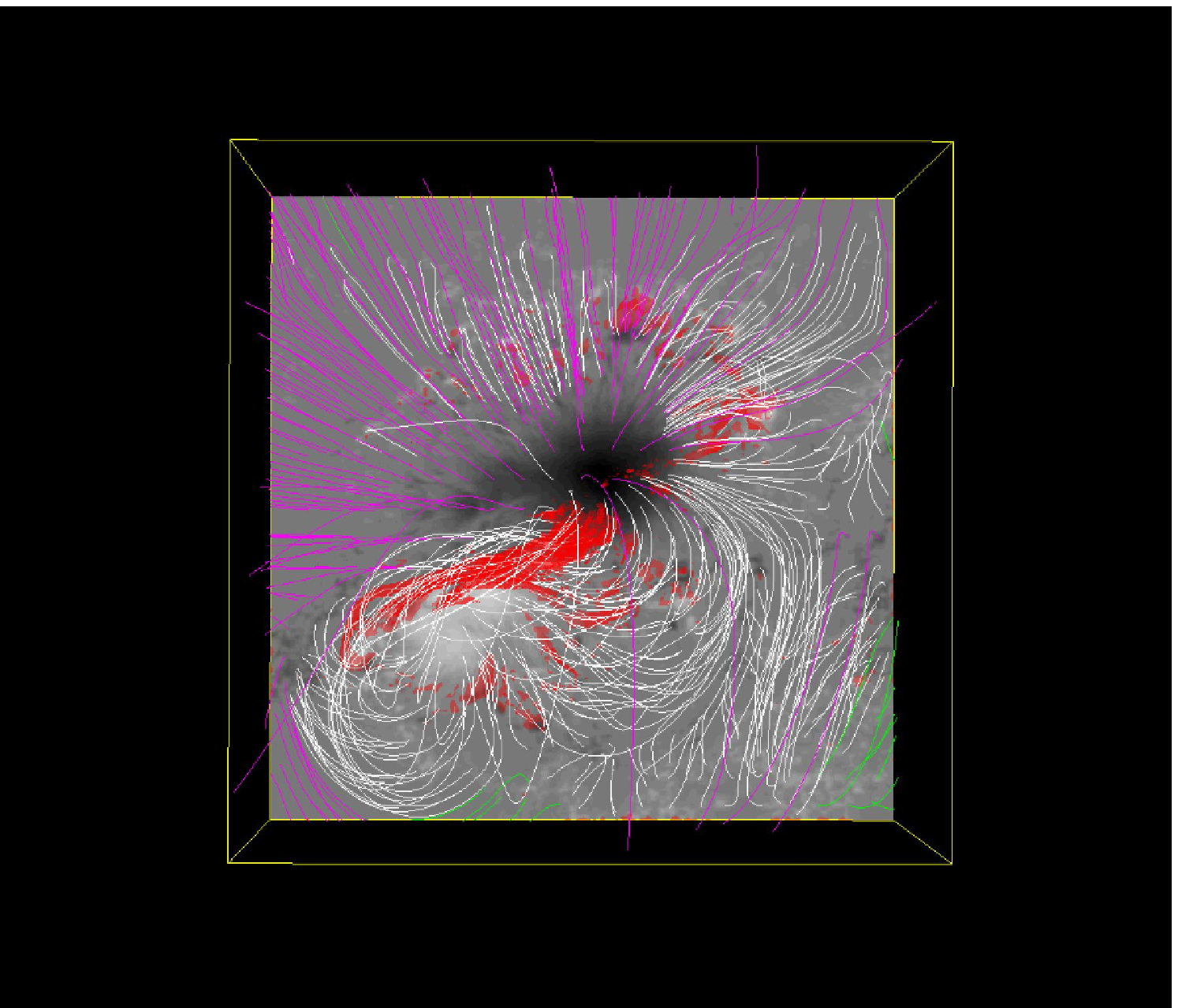,width=7.9truecm,bbllx=58bp,bblly=63bp,bburx=357bp,bbury=302bp,clip=} \psfig{figure=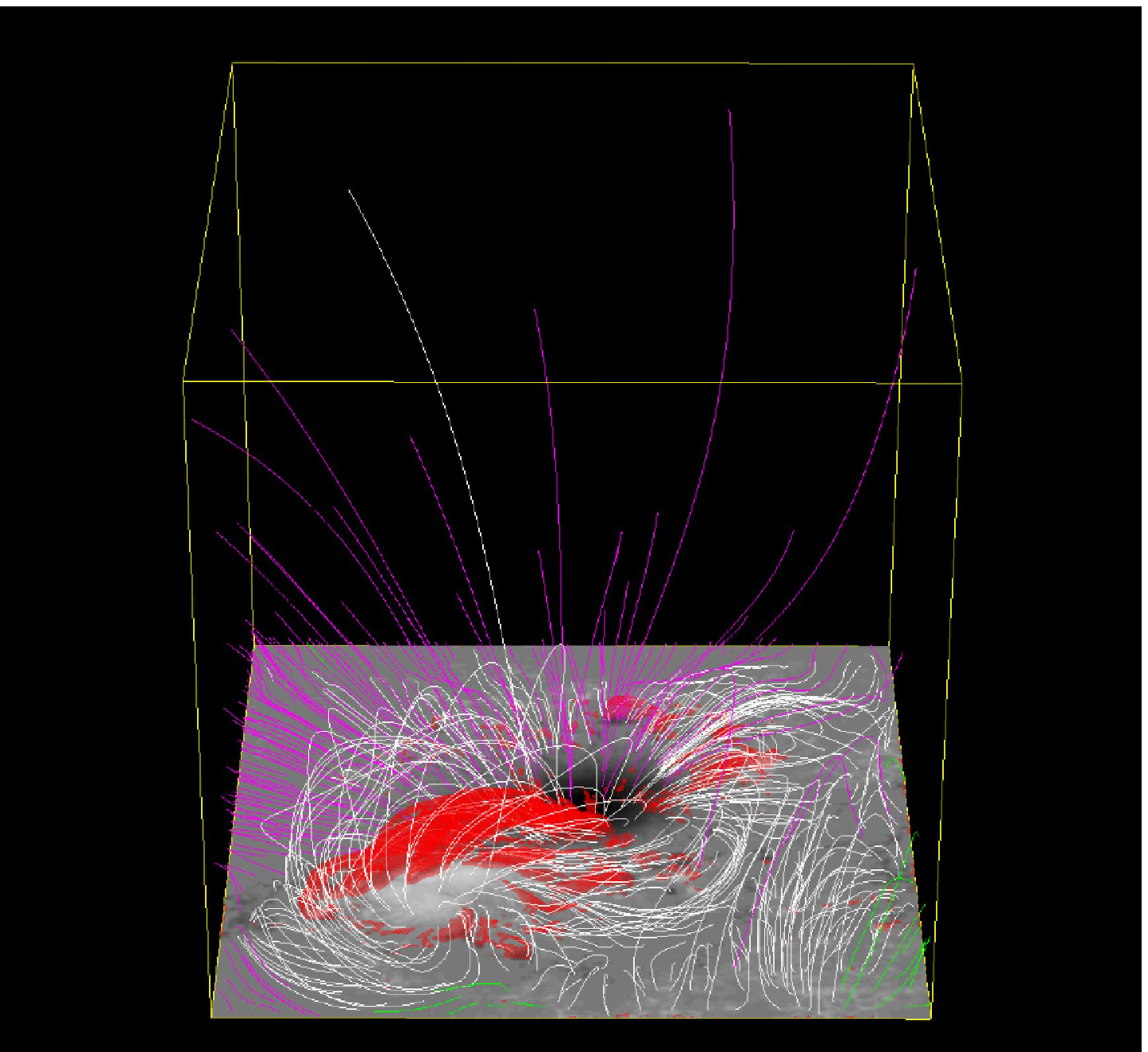,width=7.9truecm,bbllx=58bp,bblly=5bp,bburx=357bp,bbury=244bp,clip=}}
\vskip 0.1truecm
\centerline{\psfig{figure=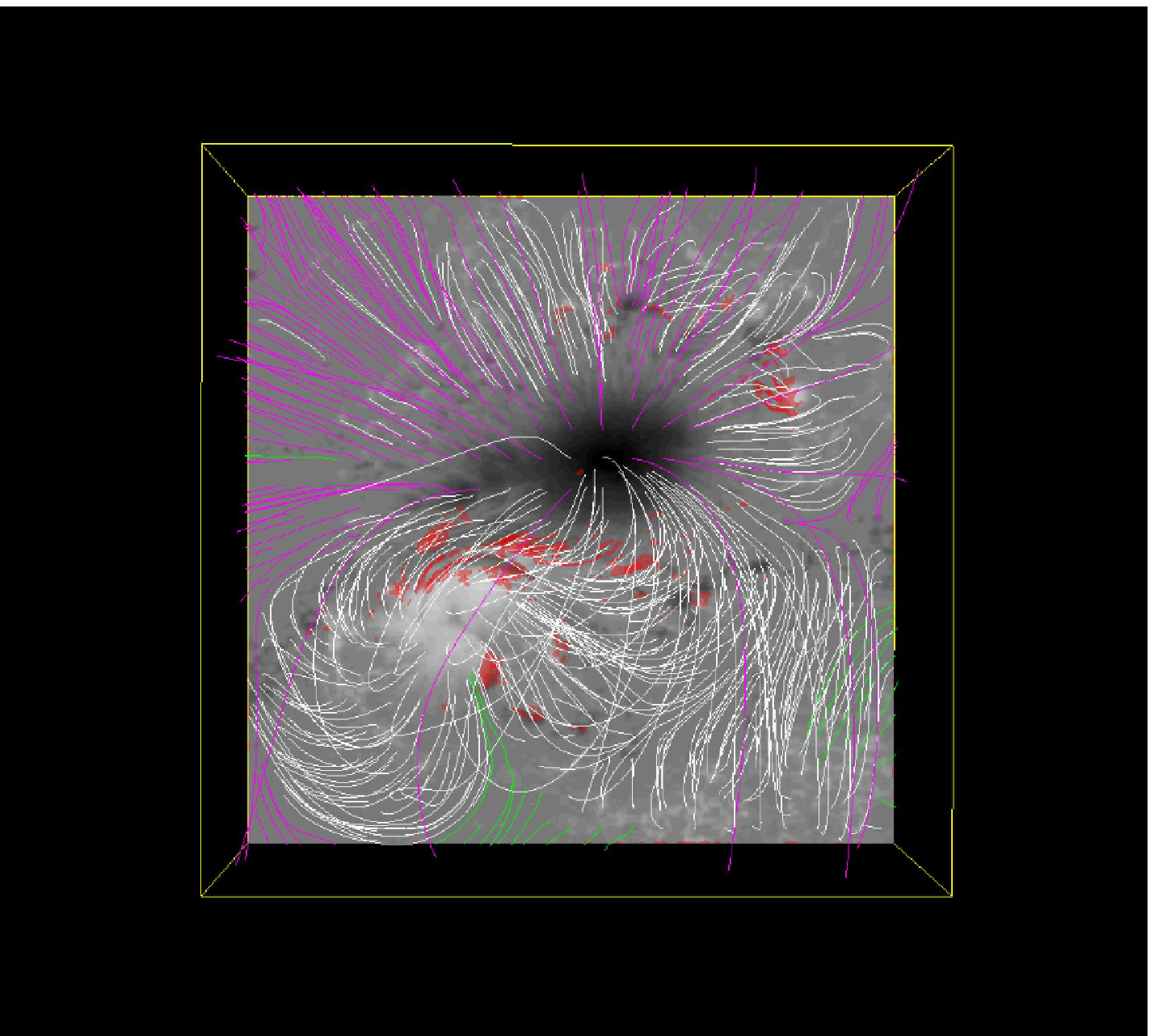,width=7.9truecm,bbllx=49bp,bblly=63bp,bburx=348bp,bbury=302bp,clip=} \psfig{figure=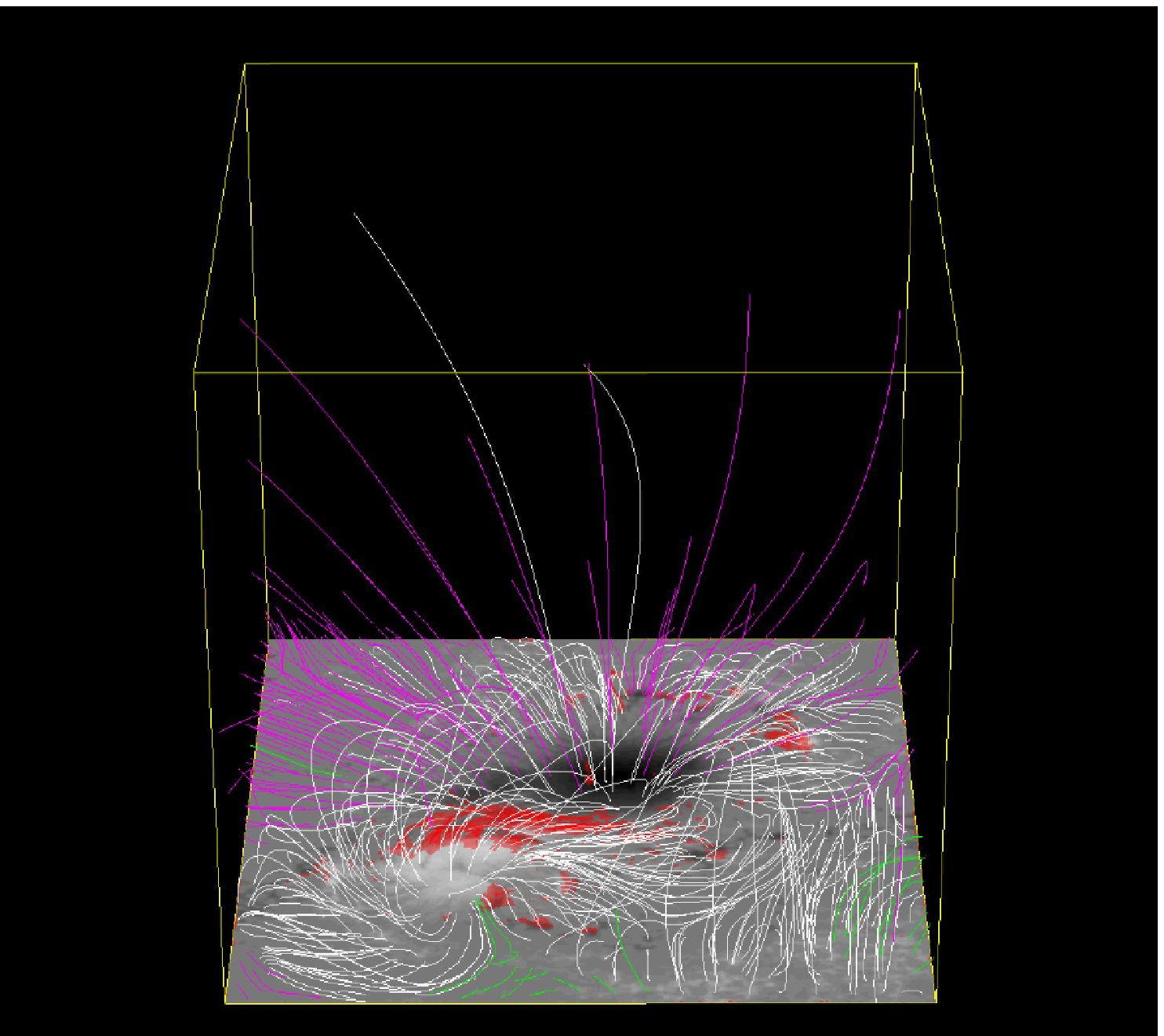,width=7.9truecm,bbllx=58bp,bblly=5bp,bburx=357bp,bbury=244bp,clip=}}
\centerline{f3[a,b,c,d].eps}
\end{figure}
\clearpage

\begin{figure}
\plotone{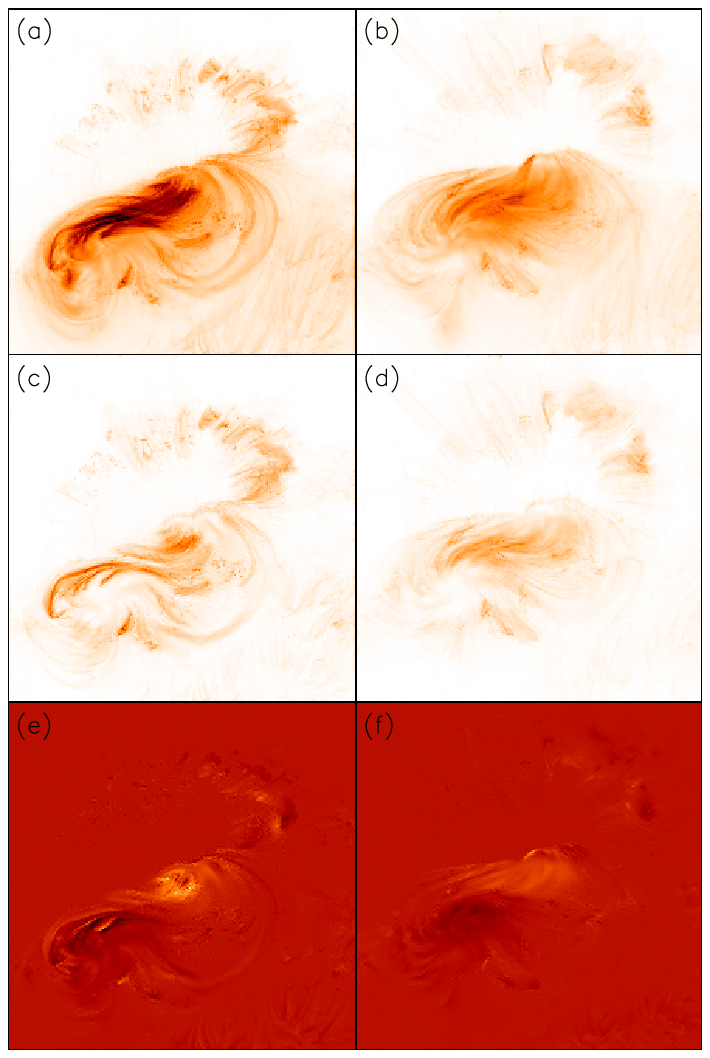}
\centerline{f4.ps}
\end{figure}

\end{document}